\documentclass{aip-cp}

\usepackage[numbers]{natbib}
\usepackage{rotating}
\usepackage{graphicx}

\begin{document}

% Title portion
\title{Upgrades for the Precision Proton Spectrometer at the LHC: Precision Timing and Tracking Detectors}

\author[lip]{Michele Gallinaro\corref{cor1}}
%\eaddress[url]{http://www.aip.org}
%\author[aff2,aff3]{Author's Name}
%\eaddress{anotherauthor@thisaddress.yyy}

\affil[lip]{
Laborat\'orio de Instrumenta\c{c}\~ao e F\'isica Experimental de Part\'iculas\\
        LIP Lisbon\\ Av. Elias Garcia, 14 - 1000-149 Lisboa - Portugal}
\corresp[cor1]{on behalf of the CMS and TOTEM collaborations}

\maketitle

\begin{abstract}
The CMS-TOTEM Precision Proton Spectrometer (CT-PPS) is an approved project to add tracking and timing information at approximately $\pm$210~m from the interaction point around the CMS detector. It is designed to operate at high luminosity with up to 50 interactions per 25~ns bunch crossing to perform measurements of e.g. the quartic gauge couplings and search for rare exclusive processes. During 2016, CT-PPS took data in normal high-luminosity proton-proton LHC collisions. In the coming years, high radiation doses and large multiple-vertex interactions will represent difficult challenges that resemble those of the high-luminosity LHC program. A coordinated effort of detector upgrades with the goal of reaching the physics goals while mitigating the degradation effects is under way. Upgrades to the tracking and timing detectors are discussed.
\end{abstract}

% Head 1
\section{INTRODUCTION}
%\vspace{-0.1cm}
The CMS-TOTEM Precision Proton Spectrometer (CT-PPS)~\cite{ctppstdr} is a joint project of the CMS and TOTEM collaborations aimed at measuring the surviving scattered protons 
during standard running conditions in regular "high-luminosity" fills. 
CT-PPS adds precision tracking and timing detectors in the very forward region on both sides of 
the CMS detector at about 210 meters from the interaction region to study central exclusive production (CEP) in proton-proton collisions.
CEP provides a unique method to access a variety of physics topics, such as -among others- new physics searches via anomalous production of W and Z boson pairs, 
high-$p_T$ jet production, and possibly the production of new resonances.

The CT-PPS detector consists of a silicon tracking system to measure the position and direction of the protons, and a set of timing counters to measure their arrival time.
This allows the reconstruction of the mass and momentum as well as of the z coordinate of the primary vertex of the centrally produced system.
%The CT-PPS is a magnetic spectrometer that uses the LHC magnets between the Interaction Point (IP) and detector stations to bend protons that have lost a small fraction of their momentum out of the beam profile their trajectories can be measured.
The protons that have lost a small fraction of their momentum are bent outside the beam profile by the LHC magnets between the Interaction Point (IP) and the detector stations, and their trajectories can be measured in the CT-PPS detector.
The detector covers an area transverse to the beam of about 4~cm$^2$ on each arm. It uses a total of 144 pixel readout chips and about 200 timing readout channels.

% Figure
\begin{figure}[h]
\centerline{\includegraphics[width=450pt]{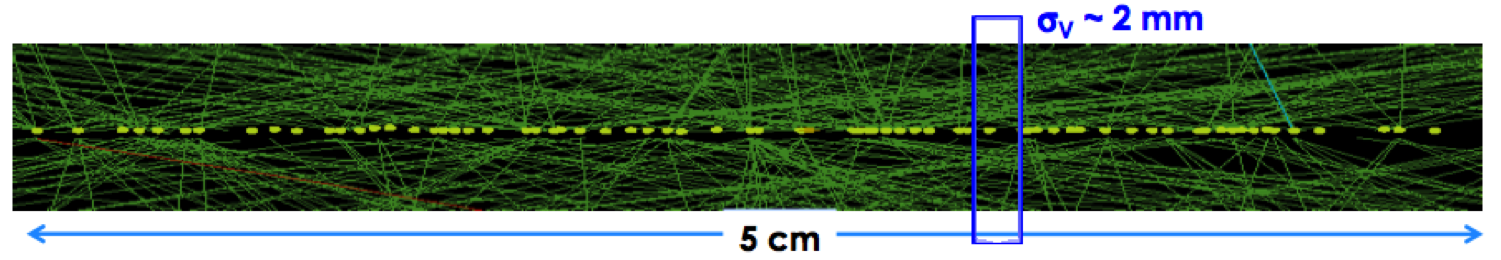}}
\caption{Zoomed-in view of an event with 78 reconstructed vertices (yellow points) and associated tracks (green lines) collected by the CMS experiment during the LHC Run~1.}
\label{figure1}
\end{figure}

Several challenges are to be met for successfully carrying out the physics program. 
%The ability to operate the detectors close to the beam (15-20$\sigma$) in order to maximize the acceptance for low-momentum loss (also known as $\xi$) protons. 
Among those is the ability to operate the detectors close to the beam center, thus enhancing the sensitivity to lower invariant mass systems. 
A close approach is limited by the interference with the beam,
%The ability to closely approach the beam is limited by reducing the interference to the beam
%impedence introduced by the detector ``pockets'', 
as well as the detector damage caused by the high radiation levels.
Studies based on radiation monitors placed on the 220~m TOTEM RP stations yield a total dose of about 100~Gray and a fluence of the order 
of $10^{12}~n_{\rm eq}$cm$^{-2}$ per 100~fb$^{-1}$ of integrated luminosity~\cite{Ravotti:2011cna}. These are values at the location of front-end electronics in the RPs. 
A proton flux of up to $5\times10^{15}$cm$^{-2}$ is expected on the sensors themselves.
%For 100~fb$^{-1}$, a proton flux of up to $5\times10^{15}$cm$^{-2}$ in the tracking sensors is expected.
%Interference to the beam impedance introduced by the beam ``pockets''
%Furthermore, the large number of pileup events
%Furthermore, the presence of multiple interactions (pileup) per bunch crossing is (both in-time and out-of-time with the collision) 
Furthermore, another challenge comes from the large background expected in the presence of multiple interactions (pileup) per bunch crossing. 
%The high-pileup environment during normal LHC collisions
The number of pileup events was approximately $\mu$=20$\div$30 during 2016, and may reach 50 or 100 in the next few years.
%, or even 200 during the high-luminosity phase of LHC.
Figure~\ref{figure1} shows an event with 78 reconstructed vertices collected by the CMS experiment during the LHC Run~1.
The dominant background results from inelastic events overlapping with two protons from single diffraction events occurring in the same bunch crossing. 
This pileup background can be reduced by using proton timing information to determine the z-position of the primary vertex.
The expected increased pileup and large radiation doses represent some of the difficult challenges that need to be addressed with a focused upgrade program.
Upgrades to the tracking and timing detector systems and associated electronics are foreseen.
These are needed in order to maintain the physics goals and be able to efficiently collect the large samples of data necessary to explore the rare processes 
that may hint at a new panorama in our current understanding of Nature.

\section{TRACKING DETECTORS}
\vspace{-0.1cm}
Silicon detectors are widely used in experiments in particle physics. The application of this detector technology is mostly suited for tracking detectors, i.e. detectors which measure the position of charged particles. The enormous progress in improving the resistance to radiation damage and the excellent spatial resolution make this technology well suited for CT-PPS.
Tracking detectors are used to measure the position and direction of the outgoing protons. This will allow to reconstruct the mass and momentum of the centrally produced system.
%Since the first use of silicon detectors in a High Energy Physics experiment, significant progress has been made.

In the past few years there has been considerable progress in silicon sensor technology.
At the LHC, both the CMS and ATLAS collaborations have been pursuing improved pixel detector design for the high-luminosity upgrades of the LHC. 
These efforts culminated with the development of "planar" and "3D" pixel sensor designs~\cite{Parker:1996dx,Mathes:2008ft}. 
Both technologies can be used for the CT-PPS detector due to their characteristics. 
These include: 1) efficient tracking close to the physical sensor edge providing single-hit resolution better than $30\mu$m, 
2) improved radiation hardness to withstand $5\times 10^{15}$~protons/cm$^2$ for an integrated luminosity of 100~fb$^{-1}$.
The 3D sensors consist of an array of columnar electrodes that penetrate through the silicon bulk perpendicularly to the surface. 
Due to its structure, the inter-electrode distance is decoupled from the sensor substrate thickness allowing to reduce the drift path of the charge carriers.
%(without decreasing the total generated charge). 
The close spacing between electrodes provides several advantages compared to the planar sensor design. 
Among others, they are: 1) lower depletion voltage, 2) faster charge collection time, and 3) reduced charge trapping probability, thus reducing the effects of radiation damage.
Thanks to their better intrinsic characteristics, 3D sensors are the preferred choice for CT-PPS.
%Due to better intrinsic characteristics than the planar sensors, the 3D sensor design is the preferred choice for CT-PPS. %~\cite{Ravera}.

\begin{figure}[h]
\centerline{\includegraphics[width=450pt]{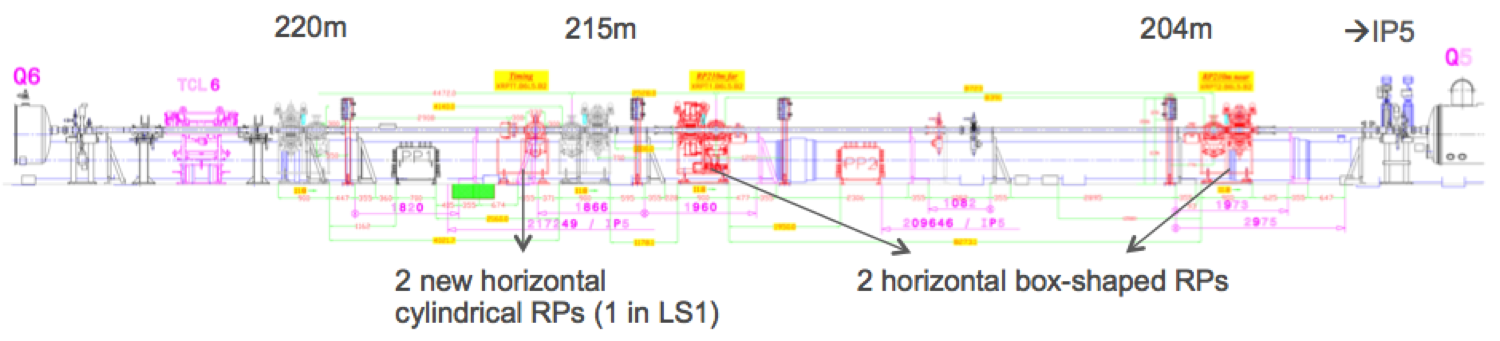}}
%\vspace{-1cm}
\caption{Layout of the beam line and the CT-PPS detectors in the 210~m region.}
\label{figure2}
\end{figure}

The CT-PPS layout (Figure~\ref{figure2}) foresees two RP stations on each side of the IP to be equipped with tracking detectors. 
%The configuration of the tracking system consists of two detector stations in each arm, for a total of four detector stations. 
These are the horizontal RPs located at $\pm$210~m. Each station will contain six planes, where each plane contains a $1.6\times2.4$~cm$^2$ pixel sensor read out by six PSI46dig readout chips (ROCs). 
%Each ROC reads 52?80 pixels with dimensions 150?100 ?m2. Given the small area of the detector, covered by a small number of individual sensors, 
%we have chosen a number of planes that provide confortable redundancy making the system resilient to possible failures. 
The design of the front-end electronics and of the DAQ is based on that developed for the ``Phase~1" upgrade of the CMS silicon pixel detectors.
The resolution of the x-coordinate is determined by the charge sharing in the pixel clusters, which depends on the detector tilt angle in the x-z plane. 
Test beam results with similar sensors indicate that for an angle of 20 degrees 
the two-pixel clusters give a resolution of the order of $10\mu$m. 
%the clusters have a resolution of the order of $10\mu$m. 
Since there is no tilt in the y-z plane, the resolution of the y-coordinate is of the order of $30 \mu$m.

%The layout of the CT-PPS foresees two RP stations on each side of the IP to be equipped with tracking detectors. 
During the 2016 data-taking period, the TOTEM silicon strip detectors were used. These detectors were not intended to sustain high radiation doses
but were used during this preliminary phase to commission the detector and DAQ operations. 
About 15~fb$^{-1}$ of data were collected with CT-PPS in "physics" mode at the safe operating position of $15\sigma$ from the beam.
Preparation of the sensors and detector packages, including the readout electronics and the associated mechanics, is ongoing. 
Starting in 2017, the 3D tracking detectors described above will be installed.
%improved silicon tracking detectors will be installed for regular CT-PPS operations.

%Preparation of the sensors and detector packages, including the readout electronics and the associated mechanics, is ongoing. 
%Installation is foreseen during the upcoming year-end technical stop to be ready  for the data-taking of 2017.

% Figure

\section{TIMING DETECTORS}
\vspace{-0.1cm}
Upgrades for the high-luminosity LHC detectors
%High-Luminosity LHC upgrades 
are driving the efforts on several fronts, including R\&D on timing detectors as an important tool to resolve multiple vertices in events with a large number of interactions.
Timing information can provide a powerful discrimination of signal and background events in the high-luminosity LHC era, 
by adding the "event time" to the "event vertex" information. 
%when the mean number of interactions per beam crossing is expected to reach$\sim$150-200. Event vertex and event time are complementary information.
In the case of CT-PPS, 
timing detectors are used to measure the time of arrival of the outgoing protons at the RP detectors on both sides of the IP in order to determine the z-coordinate of the primary vertex.
A time resolution of 10~ps corresponds to a vertex resolution of 2.1~mm. 

%Several technologies have been studied to reach this goal.

%A detector based on quartz bars and Cherenkov propagation of the light was considered earlier for good timing resolution capabilities. 
%However, after preliminary beam tests of a full module and further considerations on the detector segmentation 
%(especially close to the beam where the double-hit rate in the same bunch crossing increases) and material thickness (potentially causing multiple scattering) other technologies were considered.

There are several technologies that can be used for a fast timing detector. Considerable progress has been made on silicon detectors. 
Solid state timing detectors have the advantage of being very thin and allow for a fine channel granularity. 
Thin sensors make stacking of several layers possible, thus improving the overall time resolution. % without the worry of multiple scattering. 
%, thus improving the time resolution. 
Furthermore, small size pixels may allow a significant reduction of rate per channel.
Among key characteristics desired for good timing capabilities are large signals, fast drift, and low noise.
An increased signal brings many benefits such as the possibility of developing thin detectors with large enough signals. Thin sensors also have the advantage of collecting the electrons fast.
Furthermore, in order to improve the time resolution, large signal-to-noise ratio, small capacitance, and small leakage current are desirable.
%can be achieved by using small electrodes.
In general,
the time resolution is proportional to $\sim t_{\rm risetime}/({\rm S/N})$, where S and N are signal amplitude and noise level. 
For a risetime of $t_{\rm risetime}$$\sim $2~ns, one needs S/N$\sim$100 to get into the 20~ps regime.
However, this is only a first approximation as there are additional effects that may play a role in worsening this naive picture.
Nice reviews can be found in~\cite{Adams:2016tfm,Va'vra:2016wpv}.

Good time resolution can be achieved with mono-crystalline {\bf diamond detectors}. These devices are fast, low-noise, and radiation hard.
This is the technology chosen for CT-PPS.
%The technology chosen for CT-PPS is based on {\bf Diamond sensors}, a solid state silicon detector. 
The choice is based on the combination of several ingredients. These include: 1) fine segmentation, thus reducing the occupancy of the individual channels, 2) thin sensors, thus reducing multiple scattering, 3) improved radiation tolerance, and 4) good time resolution.
A MIP signal produces a small signal and therefore low-noise readout electronics must be used.
A single-hit time resolution of 80-100~ps was measured in beam tests~\cite{Berretti:2016zjl}. 
During 2016,  two RP stations -one on each side of the IP- were instrumented with diamond detectors. Each detector package is composed of four layers of diamond sensors.

Other technologies are being studied and extensive R\&D is pursued in parallel in order to achieve the ultimate goal of 10~ps resolution.
Extensive R\&D is performed on Avalanche Photo Diodes (APDs) for their fast timing capabilities. 
In order to achieve good timing resolution they should be operated at a high gain, but they can also work at a lower gain provided that the signal-to-noise ratio is sufficiently large.
%have been recently are an old technology, and recently 
For this purpose, 
%extensive R\&D is 
studies are performed on silicon sensors by exploiting the charge amplification in the silicon itself, 
either working in "Geiger" (as in gas RPC) or in "linear" (as in Low-Gain Avalanche Detectors, LGADs) mode. 

The progress made on LGADs has made possible to develop silicon detectors yielding large signals that are a factor of ten larger than those of traditional sensors. 
%This brings the benefit of developing thin detectors with large enough signals.
The so-called {\bf Ultra Fast Silicon Detectors (UFSDs)}~\cite{Sadrozinski:2016xxe,Cartiglia:2016voy} are LGADs optimized to achieve good timing resolution.
They can reach a gain of 10-20 for voltages of 200-240V.
They are n-on-p silicon sensors with an additional p$^+$ implant under the n-implant.
The additional p$^+$ implant generates a large field used to generate the charge multiplication with a moderate internal gain of $\sim$10.
Results of beam tests with a 180~GeV pion beam at CERN indicate 
%that UFSD with an active thickness of 45 ?m and 1.4 mm2 pad size reaches 
that a timing resolution of 35ps can be achieved for a single layer with an active thickness of $45\mu$m.
%at a bias voltage of 200 V and 26ps at 240 V.
%LGADs have therefore made possible the development of silicon sensors optimized to achieve excellent timing performance, the so called Ultra-Fast Silicon Detectors
Furthermore,
%current 
tracking devices are small and not able to provide timing information accurately, while good timing detectors are too large for an accurate position measurement.
Ultimately, UFSDs could provide precise tracking and accurate timing information at the same time.
However, charge multiplication is affected by the radiation doses expected~\cite{Kramberger:2015cga}, and more tests are foreseen to further quantify the effects.
%Further tests to evaluate the damage due to radiation are foreseen in the near future.
%
One timing layer of UFSDs for CT-PPS is formed by 32 strips; they are readout by four 8-channel amplifier-discriminator custom-made ASICs with low-power consumption.
The readout chain is based on a time-over-threshold method to measure the timing information.
Finalization of UFSD sensor production and readout electronics is under way in order to install the first detectors already in 2017.

%Avalanche Photo Diodes (APDs) are silicon sensors 
APDs with large amplification are also being exploited for their good timing capabilities. 
These devices -- also known as {\bf HyperFast Silicon Detectors (HFSD)}~\cite{White:2014oga} -- 
use a ``deep depleted'' technology and operate at gains of $\sim$500 in a depleted region deep inside the sensor. They require a large electric field with a voltage of $\sim$1.8kV.
This technology minimizes the drift region to a thin layer of a few $\mu$m near the surface while the fast avalanche develops in the depleted region, thus improving the timing performance.
It has a risetime of $\sim$1ns; sensors with an area of  $\sim$1cm$\times$1cm can also be made.
Developed by RMD (Radiation Monitoring Devices Inc.), this is a radiation-hard technology~\cite{mcclish,sebastian}.
%tested up to $\sim$10$^{14}$ neutrons/cm$^2$~\cite{mcclish,sebastian}. 
Tests are under way to further characterize the effects.
%Tests with a muon beam indicate that a time resolution of less than 20~ps can be achieved.
A time resolution of less than 20~ps was measured in tests with a muon beam.

%where large gains of $\sim 500$ are achieved in a depleted region deep inside the sensor, after a thin layer of a few $\mu$m near the surface.
%where after a thin layer of a few $\mu m$

\section{SUMMARY}
\vspace{-0.1cm}
The CMS-TOTEM Precision Proton Spectrometer (CT-PPS) was approved in 2014 to extend the coverage of the central detector 
to the region close to the beam at $\pm210$~m from the interaction region.
CT-PPS will allow -for the first time- measurements of central exclusive processes at the electroweak scale, 
and it will enhance sensitivity to quartic gauge couplings and searches for rare exclusive processes.
%After an accelerated 
After an intense preparatory phase, CT-PPS started taking "physics" data in 2016 during high-luminosity fills in proton-proton collisions.
A program of detector upgrades is under way to prepare for the regular insertion of the detectors close to the beam during the next few years, 
with the goal of collecting 100~fb$^{-1}$ of data.
The ability to withstand high radiation doses while retaining accurate tracking reconstruction and fast timing capabilities is the main challenge to meet.
%High radiation doses while retaining accurate tracking reconstruction and fast timing capabilities are the challenges to meet.
Tracking and fast timing detector upgrades have been discussed, together with the R\&D program that is actively being pursued for further detector improvements.
%upgrades for tracking and fast timing detectors have been discussed, together with the R\&D program that is actively being pursued for further detector improvements.

%Upgraded tracking and fast timing detectors are being prepared, and a program R\&D is actively pursued for further detector improvements. 
%Tracking resolution of 10$\mu$m and a time resolution of 20~ps are 

% Sections that will go in second font

% Acknowledgement
\section{ACKNOWLEDGMENTS}
\vspace{-0.1cm}
To my TOTEM and CMS colleagues who strenuously contributed to the success of this difficult project becoming a reality, without forgetting that there are still hurdles ahead. 
To the RD51 colleagues for a stimulating environment 
where curiosity thrives, something that is often forgotten in our everyday life. 
% and for reminding that curiosity is a fundamental part in our everyday life.
To the Organizers for an interesting meeting in a beautiful place, and for a kind and relaxed atmosphere.

% References

%\nocite{*}
%\bibliographystyle{aipnum-cp}%
%\bibliography{sample}%
%\bibliography{diffraction2016}%
%%
%% BIBLIOGRAPHY
%%
%\bibliography{auto_generated}

\end{document}